\documentstyle[a4,12pt,epsf]{article}
\begin{document}
\title{\bf Time of arrival \\
through interacting environments: \\
Tunneling processes}
\author{Ken-Ichi Aoki\footnote{Electronic address : 
aoki@hep.s.kanazawa-u.ac.jp}, 
~Atsushi Horikoshi\footnote{Electronic address : 
horikosi@hep.s.kanazawa-u.ac.jp}, 
~and ~Etsuko Nakamura\footnote{Electronic address : 
etsuko@hep.s.kanazawa-u.ac.jp}\\
\\
Institute for Theoretical Physics, Kanazawa University, \\
Kakuma-machi Kanazawa 920-1192, Japan}
\date{July 2000}
  \maketitle
\vspace{-100mm}
\begin{flushright}
quant-ph/9912109\\
KANAZAWA/99-14
\end{flushright}
\vspace{80mm}
%%%%%%%%%%%%%%%%%%%%%%%% Abstract %%%%%%%%%%%%%%%%%%%%%%%%%%%%% 
\begin{abstract}
 We discuss the propagation of wave packets 
 through interacting environments. 
 Such environments generally modify the dispersion 
 relation or shape of the wave function. 
 To study such effects in detail, 
 we define the distribution function $P_{X}(T)$, 
 which describes the arrival time $T$ of a packet 
 at a detector located at point $X$. We calculate $P_{X}(T)$ for
 wave packets traveling through a tunneling barrier and find   
 that our results actually explain recent experiments. 
 We compare our results with Nelson's stochastic interpretation of
 quantum mechanics and resolve a paradox previously apparent in 
 Nelson's viewpoint about the tunneling time. 
\end{abstract}
%%%%%%%%%%%%%%%%%%%%%%%% Section 1 %%%%%%%%%%%%%%%%%%%%%%%%%%%%% 
\section{Introduction}
\quad We are interested in the behavior of quantum particles, 
that is, wave packets propagating through interacting environment. 
In general, there are two types of environment. 
One is the ordinary medium (plasma, dielectric, etc.) 
which consists of ``matter'' [1-4].  
The other is the nontrivial structure of the vacuum due to field
theoretical fluctuations \cite{lpt}
or effects of quantum gravity \cite{aemns,gar}. In both cases, 
the presence of such environments will modify the dispersion 
relation of particles, $E=f(p)$, or modify the shape of the wave packet. 
Observation of the arrival time of particles through such environments is
a way to see the effects of these modifications.
Recently, these effects have been tested in two fields, astrophysics and
quantum optics.
The first is the observation of arrival times of photons from distant
astrophysical sources such as $\gamma$-ray bursters. 
Several models of quantum gravity suggest 
that the velocity of light has an effective energy dependence 
due to the modified dispersion relation 
induced by the nontrivial structure of space-time at distances 
comparable to the Planck length. 
To confirm this effect, it is necessary to observe a certain difference
of the arrival time of photons with different energies, and 
$\gamma$-ray bursters work for this purpose \cite{aemns}. 
As a result, a lower bound on the energy scale of quantum gravity 
is obtained \cite{sch}. 
The second recent test is observation of tunneling of photons.    
Chiao and co-workers constructed an elaborate stadium for the race between 
photons propagating in the vacuum and through an optical barrier, 
and measured their arrival times \cite{exp1,exp2}.
They found that the photon tunneling through the barrier arrived 
at the goal earlier than the other photon traveling in the vacuum. 
Although this result implies superluminal velocity 
of the tunneling photon, it does not mean causality violation, 
because in this case the group velocity itself 
does not transport any information at all. 
The apparent superluminality results from reshaping of wave packets 
while tunneling. 
Similar phenomena can be found in absorbing media \cite{tfi}.
Anyway, in both experiments, measurement of the arrival time 
of wave packets plays an essential role. \par 
However, 
there is no clear definition of arrival time in quantum mechanics.
This has its root in the well-known fact that time is not an operator 
but a parameter in quantum mechanics. 
Though many authors have attempted to define an operator 
of arrival time and construct its eigenstates,   
a satisfactory formulation has not yet been obtained [10-25].
%\cite{ab,dm,leav,ml,grt,gian,aopru,mlp,muga,ljpu,bspm,lf,kirpol,kw}  
In this article we define a distribution function $P_{X}(T)$, 
which describes the arrival time of packets at a detector located 
at point $X$. 
In terms of $P_{X}(T)$, we can compute a mean arrival time
$\left\langle T\right\rangle _{X}$. 
Of course we assume an ideal detector and our definition of $P_{X}(T)$
might not exactly correspond to the physical measurement process. 
However, concrete calculation of $P_{X}(T)$ shows us clearly 
the dynamical properties of propagation of packets 
through interacting environments. \par 

We investigate the arrival time distribution $P_{X}(T)$ numerically
for nonrelativistic massive particles traveling through a potential 
barrier in one space dimension, that is, tunneling processes. 
This might be a simple model for the experiment 
of Chiao and co-workers. 
In this case the existence of a potential barrier $V(x)$ 
causes reflection and transmission of packets; 
therefore the behavior of $P_{X}(T)$ will be highly nontrivial, 
depending on various parameters. 
How to deal with time in tunneling processes is also known 
as the tunneling time problem. 
The problem arises from the paradox that a particle 
under a potential greater than the particle's energy seems to move 
with a purely imaginary velocity. 
In recent developments of nanotechnology, 
the study of the tunneling time has great significance 
because it might enable us to estimate the response 
time of nanodevices \cite{npt}. 
Various approaches to the tunneling time have been proposed 
by many authors [27-36];
%\cite{w,b,h-s,l-m,aagi,yamada,bkr} 
however, it seems difficult to define it uniquely
\footnote{``The systematic projector approach'' has been proposed as a
unifying theory of the various times proposed so far \cite{muga3}.}.
Therefore we need to define effective tunneling times 
for each system and each purpose. 
We have no intention of wrestling with the general theory 
of tunneling time now; therefore, we restrict ourselves to analyzing 
the time of appearance of the packet in the exit 
of the potential barrier and how it moves after that. 
These two notions determining the arrival time difference 
have usually been confused.
In this article we will distinguish them clearly. \par 
Finally we consider the real-time stochastic interpretation 
of quantum mechanics introduced by Nelson \cite{nel}. 
Since it utilizes the real-time trajectories of quantum particles 
as sample paths, we can construct an appropriate time distribution 
from ensemble of sample paths. 
This is why Nelson's approach is expected to be effective 
for time problems in quantum mechanics. 
In particular, it is interesting to attack the tunneling time problem 
from this approach because we can trace the particle's real-time motion
even under the tunneling potential.
Actually it has been found that the tunneling particle ``hesitates'' 
in front of the barrier \cite{ohba}.  
This property seems paradoxical because it implies 
that the particle tunneling through the barrier should always 
be delayed compared with the free one due to this hesitation and it 
seems contradictory to the advancement of the peak of the wave packet
as seen in the experiment of Chiao and co-workers. 
Is it a real paradox?\par
It is clear that Nelson's approach can reproduce 
any physical quantities of the usual quantum mechanics 
by averaging them about the sample path ensemble. 
However, there is no reason that any ``observables'' classically
defined in Nelson's stochastic procedures should have corresponding
quantities in the standard quantum mechanics. 
We will compute the arrival time distribution in Nelson's approach 
and compare it with our $P_{X}(T)$. 
Then we clarify the real physical meaning of the ``hesitation'' 
and show that there is no paradox at all. Furthermore, we mention that
Nelson's interpretation can explain the characteristic behavior of
$\left\langle T\right\rangle _{X}$ for tunneling particles very well.
%%%%%%%%%%%%%%%%%%%%%%%% Section 2 %%%%%%%%%%%%%%%%%%%%%%%%%%%%% 
\section{Definition of the arrival time distribution}
\quad First we will briefly review previous attempts 
to define a time of arrival operator and their difficulties. 
In the 1960s, Aharonov and Bohm quantized the representation 
of the classical arrival time for the free particle 
at a point $X=0$ \cite{ab},
\begin{eqnarray}
 T=-m\frac{x}{p} ~~\rightarrow ~~\hat{T}=-\frac{m}{2}
\left(\hat{x}\frac{1}{\hat{p}}+\frac{1}{\hat{p}}\hat{x}\right).\label{(1)}
\end{eqnarray}
Here $x$ and $p$ are the initial position and momentum, respectively, 
where we work in the Heisenberg picture. Because $\hat{T}$ satisfies 
$\left[\hat{T},\hat{H}\right]=i\hbar$, it seems a good definition.
We construct its eigenstates  
$\hat{T}\left|T\right\rangle =T\left|T\right\rangle$.
\footnote{In order to obtain a complete set, one needs two
eigenstates $\left|T,\pm\right\rangle$ for every value of $T$
\cite{muga}.}
However, these eigenstates turn out to be not orthogonal,
\begin{eqnarray}
 \left\langle p|T\right\rangle &\propto&[\theta(p)+i\theta(-p)]
 \sqrt{p}e^{ip^2T/2m\hbar},\label{(2)} \\
 \left\langle T|T'\right\rangle &\propto&\delta(T-T')
 -\frac{i}{\pi}{\rm P}\frac{1}{T-T'},\label{(3)} 
\end{eqnarray}
where P represents Cauchy's principal value. 
That is, $\hat{T}$ is not Hermitian. 
The origin of difficulty is the singular behavior of $\hat{T}$ at $p=0$. 
Recently the regularization of $\hat{T}$ with an infrared momentum
cut off \cite{grt} and an interpretation by means of
the positive-operator-valued measure were proposed \cite{gian}. 
However, the validity of this procedure is not clear \cite{aopru,mlp}. 
In the first place, there is no one-to-one correspondence 
between the operator representation in quantum theory
and the classical representation, and it becomes more complicated 
for interacting cases [19-22].
%\cite{muga,ljpu,bspm} 
\par
Now we will not insist on defining an arrival time operator; 
rather, we try to construct an arrival time distribution directly.
We suppose that there is a detector on the path along the motion 
of wave packets and it counts the particle according to the value 
of the wave function $\psi (X,t)$ at every time $t=T$. 
Supposing the detector is ideal, we directly define 
{\it the arrival time distribution} $P_{X}(T)$ from $\psi (X,T)$,
\begin{eqnarray}
 &&P_{X}(T)dT=\frac{\rho_{X}(T)dT}{\displaystyle
 \int_0^{\infty}dT \rho_{X}(T)},\quad
\rho_{X}(T)=\left|\psi(X,T)\right|^2. \label{(4)} 
\end{eqnarray}
\par
Although Eq. (\ref{(4)}) looks like a trivial definition in our picture,
we will derive it, clarifying our system setup and assumptions.
We consider a system consisting of a particle and a detector 
located at $x=X$. 
If there is no interaction between them, the system Hamiltonian $H_0$ 
and the system state $\left |\Psi\right\rangle$ are given by  
\begin{eqnarray}
H_0&=&H_{\rm p}\otimes {\bf 1}+{\bf 1}\otimes H_{\rm D}, \label{(a5)}\\
\left |\Psi\right\rangle &=&
\left |\psi\right\rangle\otimes\left |D\right\rangle ,\label{(a6)}
\end{eqnarray}
where $H_{\rm p}$ is the particle Hamiltonian, 
$\left |\psi\right\rangle$ is the particle state, 
and similarly $H_{\rm D}$ and $\left |D\right\rangle$ 
are those of the detector. 
We define the total Hamiltonian $H$ by adding the interaction
Hamiltonian $H_{\rm I}$ between the particle and the detector,
\begin{equation}
H=H_0+H_{\rm I},\quad H_{\rm I}=gV_{\rm p}(x)\otimes V_{\rm D}.
\label{(a7)}
\end{equation}
For simplicity, we consider a detector 
whose state consists essentially of two components, 
\begin{eqnarray}
 \left |\downarrow\right\rangle
 =\left(\begin{array}{c}
         0\\
         1
	\end{array}\right),{\rm unreacted},
 \qquad
 \left |\uparrow\right\rangle
  =\left(\begin{array}{c}
	  1\\
          0
	 \end{array}\right),{\rm reacted}. \label{(a8)}
\end{eqnarray}
Corresponding to this representation, we set the interaction potentials
\begin{eqnarray}
 V_{\rm p}(x)=\delta (x-X), \label{(a9)}\\
 V_{\rm D}=\left(\begin{array}{cc}
		 0 & 1\\
                 1 & 0         
	        \end{array}\right), \label{(a10)}
\end{eqnarray}
which induces a transition $\left |\downarrow\right\rangle \Rightarrow 
V_{\rm D}\left |\downarrow\right\rangle=\left |\uparrow
\right\rangle$. 
This choice of $V_D$ should be meaningful only 
in the first order of $g$.
\par
Now we consider the time evolution of the system from $t=0$ to $T$. 
We prepare the initial state
$\left |D(0)\right\rangle=\left |\downarrow\right\rangle$ 
and evaluate a quantity $R_X(T)$, which is the probability 
that the state $\left |D(t)\right\rangle$ is found 
to be $\left |\uparrow\right\rangle$ when $t=T$. 
From $R_X(T)$, we get $P_X(T)\Delta T$, 
which is the probability that the transition 
$\left |\downarrow\right\rangle \Rightarrow 
 \left |\uparrow\right\rangle$ 
occurs in a time interval $[T,T+\Delta T]$, that is,
\begin{equation}
P_X(T)\Delta T=R_X(T+\Delta T)-R_X(T)
\stackrel{\rm \Delta T\to 0}{\sim}P_X(T)dT=dR_X(T).\label{(a11)} 
\end{equation}
\par Next we evaluate $R_X(T)$ in terms 
of the particle wave function $\psi(x,t)$.
We now assume that the detector reacts only once incoherently, 
and therefore we calculate only in the first order of $g$. 
Adopting the interaction picture, the time evolution of the
state can be represented as follows in the first order of $g$:  
\begin{eqnarray}
 \left |\Psi (T)\right\rangle _{\rm I}
 &=&{\rm T}e^{-i/\hbar \int_0^{T}\!dt\,gV_{\rm pI}(x,t)
              \otimes V_{\rm DI}(t)}
    \left |\psi (0)\right\rangle _{\rm I}
    \otimes\left |D(0)\right\rangle _{\rm I} \nonumber\\
 &\simeq&\left |\psi (0)\right\rangle _{\rm I}
         \otimes\left |D(0)\right\rangle _{\rm I} \nonumber \\
 &&-\frac{i}{\hbar}\int_0^{T}\!dt\,g
                   V_{\rm pI}(x,t)\left |\psi (0)\right\rangle _{\rm I}
                   \otimes V_{\rm DI}(t)\left |D (0)\right\rangle _{\rm I}
 \nonumber \\
 &\equiv& \left |\Psi (0)\right\rangle _{\rm I}
         +\overline{\left |\Psi (T)\right\rangle _{\rm I}}
 ,\label{(a14)}
\end{eqnarray}
where T represents the time ordered product.  
$\left |\Psi (0)\right\rangle _{\rm I}$ is the undetected state 
and $\overline{\left |\Psi (T)\right\rangle _{\rm I}}$ 
is the detected state, which is written in the Schr\"odinger picture as
\begin{eqnarray}
 \overline{\left |\Psi (T)\right\rangle}
 &=&-\frac{i}{\hbar}
      \int_0^{T}\!dt\,\left[e^{-iH_{\rm p}(T-t)/\hbar}gV_{\!\rm p}(x)
                            e^{-iH_{\rm p}t/\hbar}\right]
                    \left |\psi (0)\right\rangle \nonumber \\
     &&~\qquad\quad\otimes\left[e^{-iH_{\rm D}(T-t)/\hbar}V_{\rm D}
                                 e^{-iH_{\rm D}t/\hbar}\right]
     \left |D(0)\right\rangle \nonumber\\
 &\equiv&-\frac{i}{\hbar}
           \int_0^{T}\!dt\,\left |\psi (T;t)\right\rangle
                         \otimes\left |D(T;t)\right\rangle,\label{(a21)}
\end{eqnarray}
where we introduced  
\begin{eqnarray}
 \left |\psi (T;t)\right\rangle
 &\equiv&\left[e^{-iH_{\rm p}(T-t)/\hbar}gV_{{\rm p}}(x)
               e^{-iH_{\rm p}t/\hbar}\right]
         \left |\psi (0)\right\rangle, \label{(a22)}\\
 \left |D(T;t)\right\rangle
 &\equiv&\left[e^{-iH_{\rm D}(T-t)/\hbar}V_{{\rm D}}
               e^{-iH_{\rm D}t/\hbar}\right]
         \left |D(0)\right\rangle. \label{(a23)}  
\end{eqnarray} 
We obtain $R_X(T)$ in terms of the norm of the detected state,
\begin{equation}
R_X(T)=\overline{\left\langle\Psi (T)\right.}
 |\overline{\left .\Psi (T)\right\rangle},\label{(a24)} 
\end{equation}
under our approximation of weak coupling.
Now we apply a macroscopic decoherence condition,
\begin{equation}
 \left\langle D(T;t_1)\right.|\left .D(T;t_2)\right\rangle
 =\delta (t_1-t_2).\label{(a25)} 
\end{equation}
This means that the states reacted at different times are orthogonal 
to each other, that is, once the detection process occurs, 
the total state effectively loses its coherence 
and looks like a mixed state. 
Of course it is not possible to satisfy this condition by working
in the two-dimensional Hilbert space in Eq. (\ref{(a8)}). 
We should describe the detector by means of an
infinite-dimensional Hilbert space to realize decoherence 
effectively \cite{hall}. 
However, we can avoid this assumption by
``switching on'' the interaction Hamiltonian during different small time
intervals in repeated experiments, instead of using the finite time
interval $[0,T]$ and differentiating with respect to $T$.\par

Under this condition, 
the evaluation of $R_X(T)$ and $P_X(T)$ is straightforward as follows: 
\begin{eqnarray}
 R_X(T)=\overline{\left\langle\Psi (T)\right.}
 |\overline{\left .\Psi (T)\right\rangle}
 &=&\frac{1}{\hbar ^2}\int_0^{T}\!dt_1\int_0^{T}\!dt_2\,
    \left\langle\psi (T;t_1)|\psi(T;t_2)\right\rangle
    \delta (t_1-t_2) \nonumber\\
 &=&\frac{1}{\hbar ^2}\int_0^{T}\!dt\,
    \left\langle\psi (T;t)|\psi(T;t)\right\rangle, \label{(a26)}
\end{eqnarray}
\begin{eqnarray}
 P_X(T)=\frac{\partial}{\partial T}R_X(T)
      &=&\frac{\partial}{\partial T}
               \overline{\left\langle\Psi (T)\right.}
               |\overline{\left .\Psi (T)\right\rangle} 
       =\frac{1}{\hbar ^2}
         \left\langle\psi (T;T)|\psi(T;T)\right\rangle \nonumber \\
      &=&\frac{g^2}{\hbar ^2}
         \left\langle\psi (0)\left|e^{iH_{\rm p}T/\hbar}
                                   V^{\dagger}_{\rm p}(x)V_{\rm p}(x)
                                   e^{-iH_{\rm p}T/\hbar}\right|\psi(0)
         \right\rangle \nonumber \\
      &=&\frac{g^2}{\hbar ^2}\delta (0)\left|\psi (X,T)\right|^2
 ,\label{(a27)}
\end{eqnarray}
where at the last step we inserted the complete set 
($\int dx \left|x\right\rangle \left\langle x\right |$) three times.  
Although the divergent $\delta (0)$ seems to break the validity 
of our formulation, 
we can remove this singularity by replacing the $\delta$ function 
in Eq. ({\ref{(a9)}}) with a smeared function.
We normalize the right hand side of Eq. ({\ref{(a27)}}) 
to get our expression for the arrival time distribution $P_{X}(T)$.
Using $P_{X}(T)$ we define {\it the mean arrival time} 
$\left\langle T\right\rangle _{X}$,
\begin{equation}
 \left\langle T\right\rangle _{X}=\int _0^{\infty}T P_{X}(T)dT
 .\label{(5)} 
\end{equation}
Because $P_{X}(T)$ and $\left\langle T\right\rangle _{X}$ 
have simple and general expressions, 
we can calculate them easily even for interacting cases.
Our $P_X(T)$ is often called the ``presence time distribution''
because of its behavior in the classical limit \cite{muga1}.
In order to avoid confusion, we should make clear that
the distribution $P_X(T)$ may not be interpreted as
the probability distribution of a quantum mechanical time observable. 
It is an effective distribution describing a ``relative probability.''
\par
Of course, our definition of arrival time distribution Eq. ({\ref{(4)}}) is 
not a unique one. Considering a different system setup, some people
have proposed a definition using the current $J_{X}(T)$ 
instead of $\rho_{X}(T)=\left|\psi\right|^2$ \cite{dm},
\begin{eqnarray}
 &&P_{X}^{c}(T)dT=\frac{J_{X}(T)dT}{\displaystyle
 \int_0^{\infty}dT J_{X}(T)},\quad
 J_{X}(T)=\frac{\hbar}{m}{\rm Im}\left.\left(\psi^{*}\frac{\partial
 \psi}{\partial x}\right)\right|_{x=X}.
 \label{(4J)} 
\end{eqnarray}
This definition has the serious problem 
that $J_{X}(T)$ can be negative in some cases, 
for example, detection before the potential barrier. 
Therefore we cannot identify $P_{X}^{c}(T)$ as a probability distribution.
As for detection beyond the potential barrier as we discuss below, 
$J_{X}(T)$ might effectively maintain positivity and
actually the behavior of $P_{X}^{c}(T)$ is found to be similar to ours.     
%%%%%%%%%%%%%%%%%%%%%%%% Section 3 %%%%%%%%%%%%%%%%%%%%%%%%%%%%% 
\section{Calculation of $P_X(T)$ for tunneling particles}
\quad Now let us calculate $P_{X}(T)$ and $\left\langle
T\right\rangle _{X}$ for nonrelativistic massive particles traveling
through a potential barrier $V(x)$ in one dimension. 
This is a simple model of tunneling processes
such as the experiment of Chiao and co-workers . 
Solving the time dependent Schr\"odinger equation with some initial
conditions, we can get $\psi(x,t)$.  
Except for the free case it is difficult to solve the partial
differential equation analytically, and therefore we solve it numerically.
We now employ a discretization scheme known as the Crank-Nicholson method, 
which conserves the norm of $\psi(x,t)$ even with a finite
discrete time step \cite{cal3}.
We work with the units $m=\hbar=1$ and for the initial
condition we prepare a Gaussian wave packet,
\begin{equation}
 \psi(x,0)=\left(\frac{1}{\pi\sigma^2}\right)^{1/4}
             e^{-(x-x_0)^2/2\sigma^2}e^{ik_0(x-x_0)}\label{(6)},
\end{equation}
whose mean energy is $\langle E\,\rangle=k_0^{\,2}/2+1/4\sigma ^2$, 
and we set a time independent square potential barrier $V(x)$ 
on a section $[0,d]$. 
For simplicity, in this article we work with a unique initial packet.
We fix the central wave number $k_0=2$ and in this unit we set
the width of the initial packet in the configuration space 
$\sigma =10(2/k_0)$ and the center of initial packet $x_0=-50(2/k_0)$. 
All quantities that have a dimension of time are measured 
by the $(4/k_0^2)$ unit.
Hereafter we will omit the units of the numerical values.
We change two parameters of the barrier potential $V(x)$: 
the width $d$ and the height $h$, and also the detector location $X$. 
\par
\begin{figure}[htb]  
\hspace{0mm}
 \parbox{70mm}{
 \epsfxsize=70mm     
 \epsfysize=70mm
  \leavevmode
\epsfbox{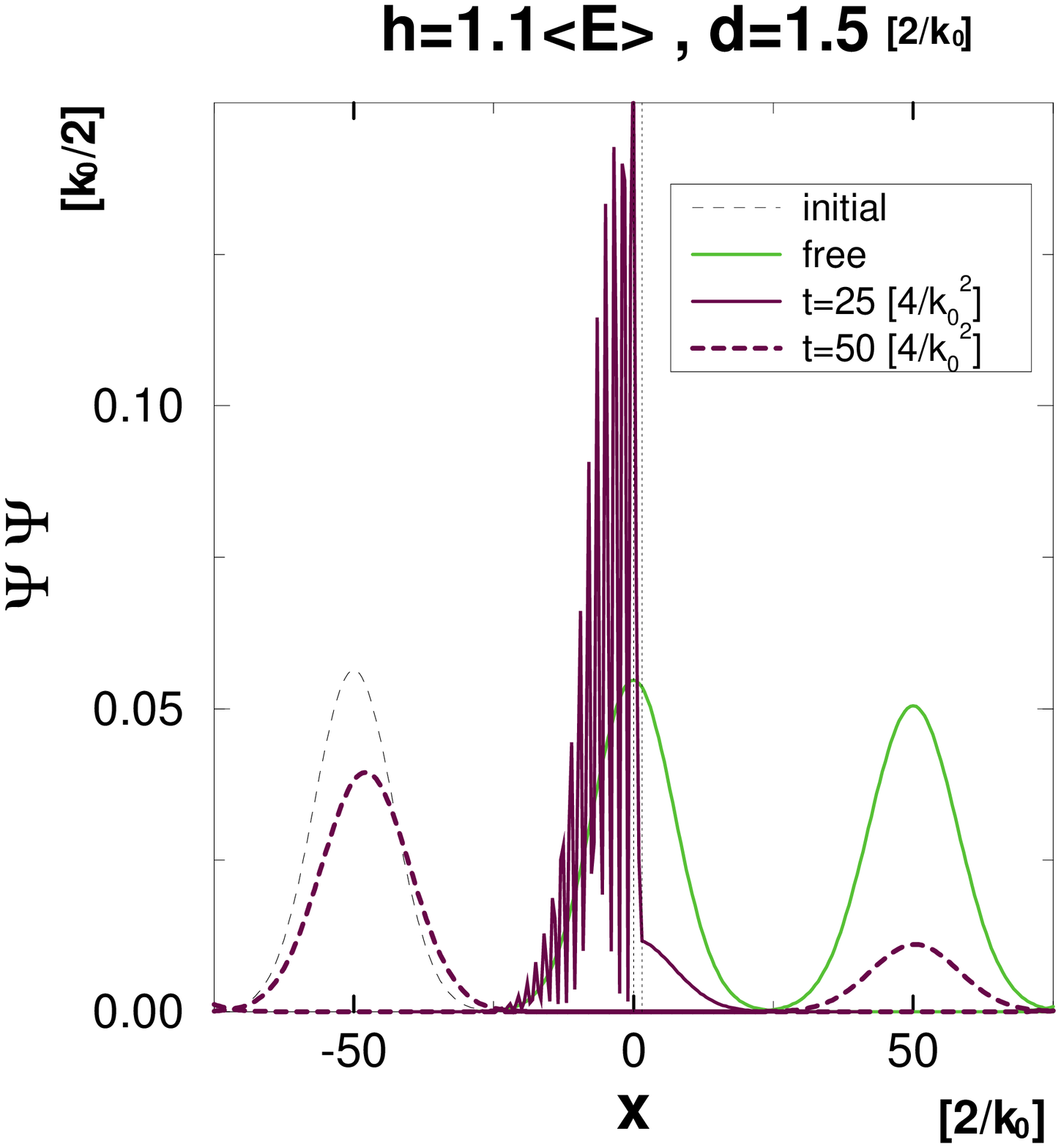}
\vspace{-5mm}
 \caption{Snapshots of the wave function squared at various times}
 \label{fig:tunnel}     
 }
\hspace{2mm} 
\parbox{70mm}{
 \epsfxsize=70mm      
 \epsfysize=70mm
 \leavevmode
\epsfbox{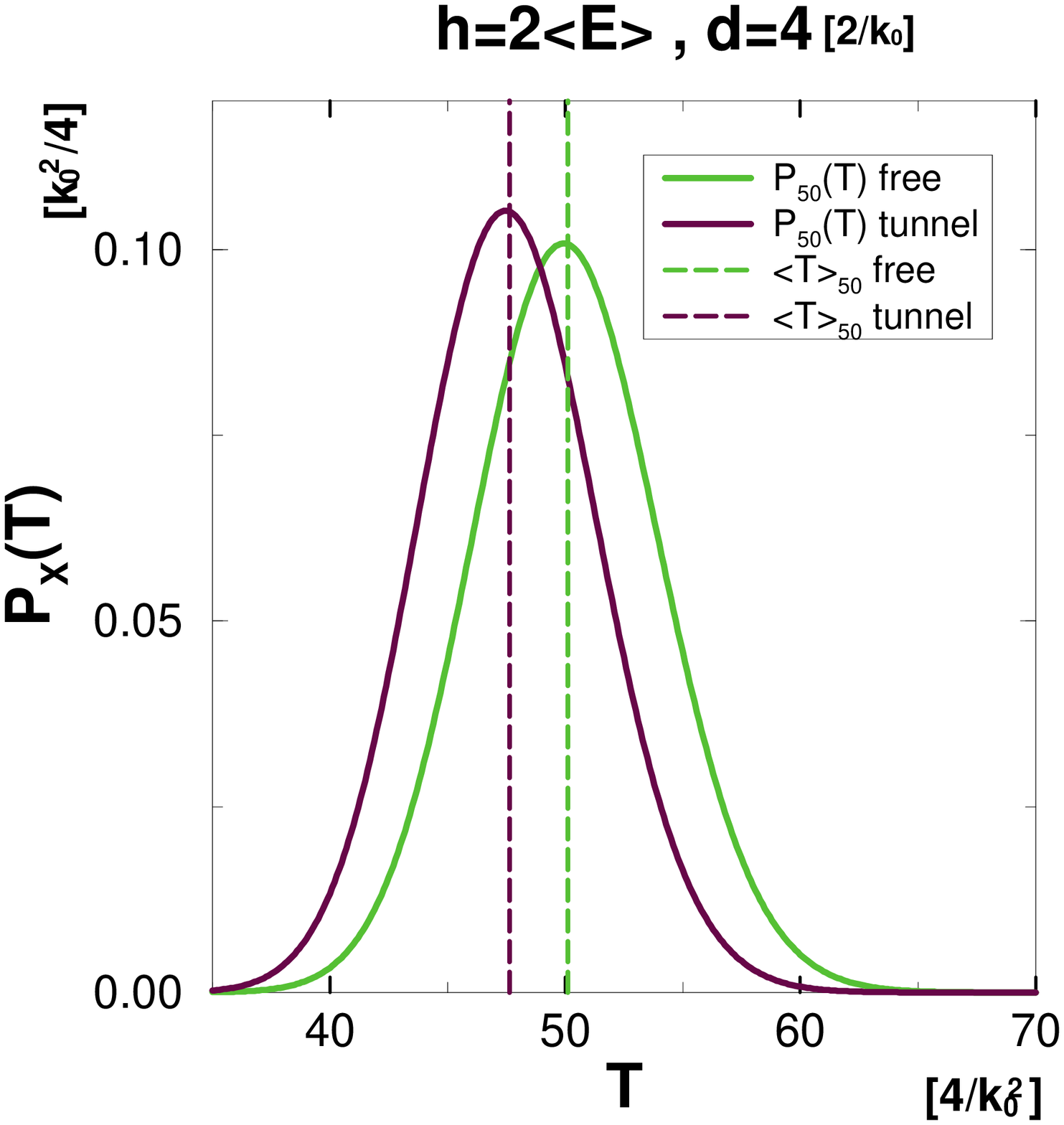}
\vspace{-5mm}
 \caption{The arrival time distribution for detection at $X=50 [2/k_0]$}
 \label{fig:detect1} 
 }
\end{figure}
Let us begin by watching the motion of wave packets
with an $h=1.1\langle E\,\rangle$, $d=1.5$ potential.
The snapshots of the motion are shown in Fig. \ref{fig:tunnel}, 
in which the free packet motion is also shown for comparison. 
Both packets spread due to the dispersive properties 
that come from their own masses. 
The packet moving through the potential barrier experiences reflection 
and transmission and the peak of the transmitted part will often advance
compared to the free packet. 
It is usually explained that this is because the higher momentum
components of the packet preferably go through the barrier and they
propagate faster than the lower momentum parts due to their dispersive
properties. 
That is, the advancement results from reshaping of the transmitted packet.
However, tracing the peak of the packet is often difficult 
because near the barrier the peak cannot be clearly identified. 
Therefore we must use more well-defined quantities,  
$P_{X}(T)$ and $\left\langle T\right\rangle _{X}$. 
\par

\begin{quote}
 {\bf Analysis 1: Detection at $X=50$} 
\end{quote} 
\quad 
Now let us calculate $P_{X}(T)$ and $\left\langle T\right\rangle _{X}$
at $X=50$ with an $h=2\langle E\,\rangle$, $d=4$ potential.
In Fig. \ref{fig:detect1}, the arrival time distributions $P_{X}(T)$ 
are plotted for the free and tunneling particles 
and the mean arrival times $\left\langle T\right\rangle _{X}$ 
are shown by dashed lines.  
The remarkable feature of $P_{X}(T)$ is the stretched tail 
and the shift of the peak caused by spread of the packet. 
For the free case, $\left\langle T\right\rangle _{X}=50.13$ is later than 
$T=50$, which is expected from the group velocity of the free packet, 
and ``the peak of $P_{X}(T)$'' $=49.94$ is earlier than $T=50$. 
It is also seen that, because of the packet's reshaping, $P_{X}(T)$ 
for the tunneling particle has a narrower shape than the free one 
and ``$\left\langle T\right\rangle _{X}$ for the tunneling particle'' 
$=47.65$ is earlier than the free one. 
However, it should be noted that only one detection far from the barrier 
cannot describe the dynamics of packets since we should discriminate
effects in and out of the potential barrier. 
Therefore we investigate detection at various points.
\begin{quote}
 {\bf Analysis 2: Detection at various points} 
\end{quote}
\quad 
When we try to give a definite answer to the so-called 
tunneling time problem, we might have to calculate the difference 
$\Delta =\left\langle T\right\rangle _{d}-\left\langle
T\right\rangle _{0}$, since this problem demands that we answer the
question ``How long does it take for the particle to tunnel across the
barrier?'' However the difference $\Delta$ does not make much sense 
because, as we can see in Fig. \ref{fig:tunnel}, the shape 
of the packet is oscillating frequently at the entrance of the barrier 
and it is difficult to distinguish between the tunneling packet 
and the reflected one, that is, 
$\left\langle T\right\rangle _{0}$ is not a good physical quantity. 
On the other hand, 
the packet has a relatively clear shape at the exit of the barrier.
Therefore we can analyze what time the packet will appear at 
the exit of the barrier and how it moves after that. \par
\vspace{5mm}
\begin{figure}[htb]  
\hspace{0mm}
 \parbox{70mm}{
 \epsfxsize=70mm     
 \epsfysize=70mm
  \leavevmode
\epsfbox{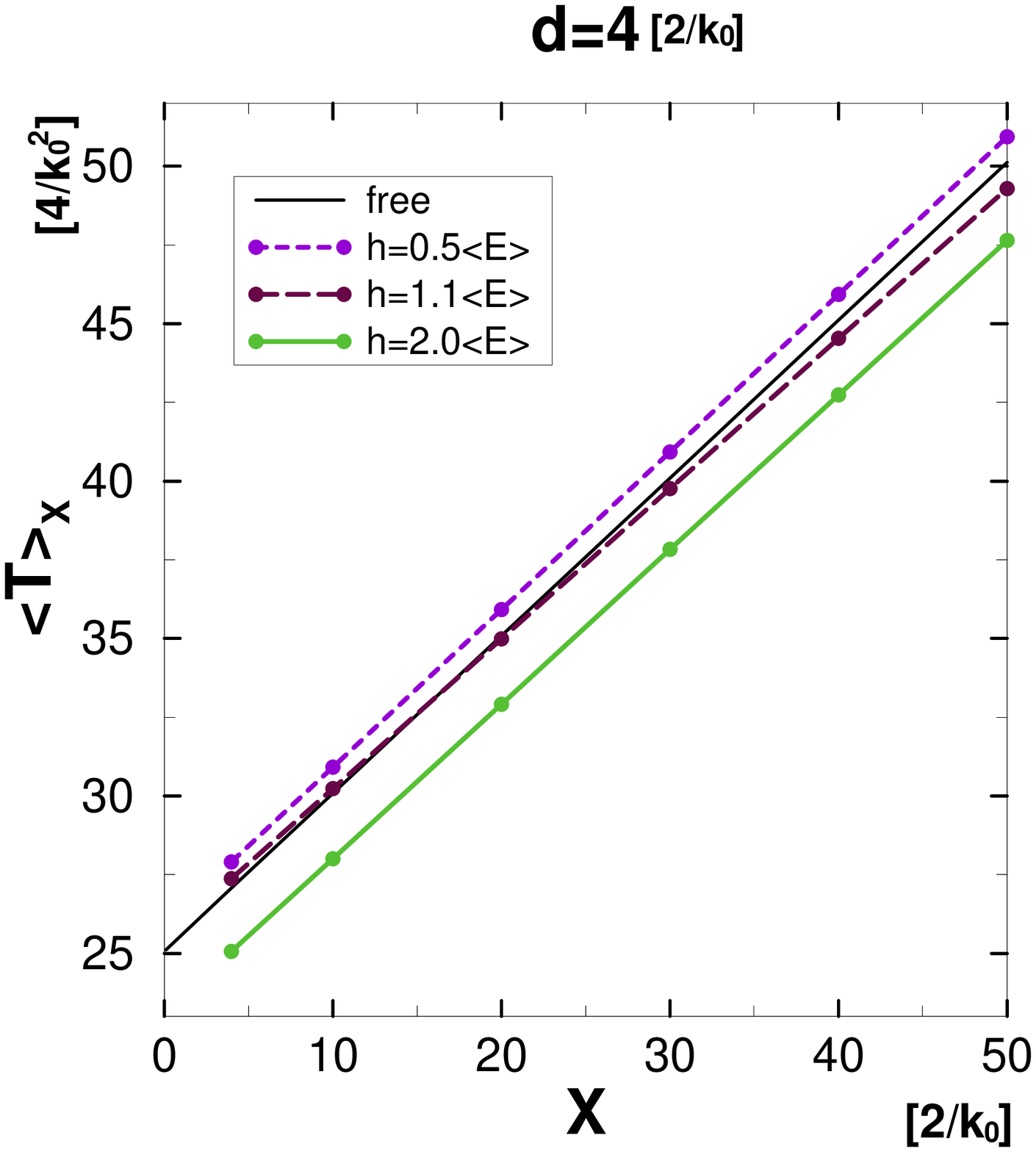}
\vspace{-5mm}
 \caption{The mean arrival time for detection at various points }
 \label{fig:avet}     
 }
\hspace{2mm} 
\parbox{70mm}{
 \epsfxsize=70mm      
 \epsfysize=70mm
 \leavevmode
\epsfbox{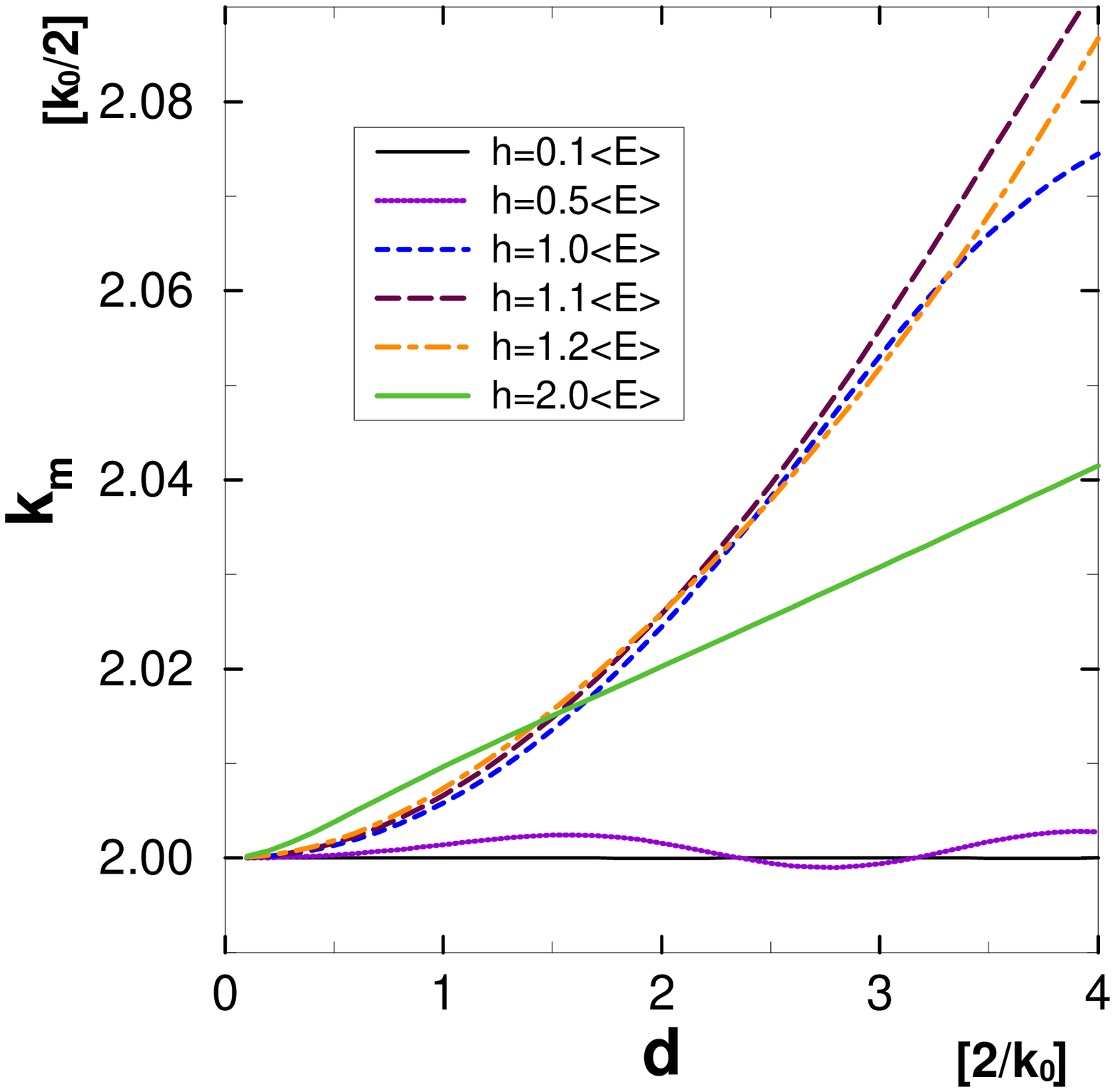}
\vspace{-5mm}
 \caption{The mean momentum of the transmitted packet}
 \label{fig:km} 
 }
\end{figure}
We calculate $\left\langle T\right\rangle _{X}$ 
for the exit of the barrier and several points after that: 
$\left\langle T\right\rangle _{X}$ at $X=d,10,20,30,40,50$, with $d=4$, 
$h=0.5\langle E\,\rangle,1.1\langle E\,\rangle,2\langle E\,\rangle$
barrier potentials (Fig. \ref{fig:avet}). 
We can see two remarkable features in this figure. 
The first is that for high barriers $\left\langle T\right\rangle _{d}$ 
is earlier than in the free case, but it is later for low barriers. 
That is, it seems that the transmitted packet arrives at the barrier 
exit earlier than the free one for tunneling dominated cases. 
These are regarded as effects in the barrier.
The second feature is that, after passing the barrier, 
the tunneling packet moves with a constant mean velocity larger than 
that of the corresponding free packet. 
This is an effect outside the barrier. 
These two types of effect are combined to cause nontrivial behaviors 
of the arrival time. For example, in the case of $d=4$, 
$h=0.5\langle E\,\rangle$, the tunneling packet arrives 
at the barrier exit $X=4$ later than the free packet; 
however, after exiting the barrier, the tunneling packet 
catches up with the free one and overtakes it at $X\simeq 15$. 
After all, it depends on $X$ which arrives at $X$ earlier, 
the tunneling or the free packet.\par
We can see the second effect clearly in the Fourier transformed
form of the transmitted packet \cite{aagi},
\begin{eqnarray}
\psi(x,t) = (4\pi\sigma ^2)^{1/4}\int \frac{dk}{2\pi}
e^{-\sigma ^2(k-k_0)^2/2}|T_k| e^{i\theta} e^{i[k(x-x_0)-\omega
t]},\label{(27)}
\end{eqnarray}
where $T_k$ is the transmission amplitude and $\theta$ is the phase. 
Using an analytically obtained $|T_k|$, the mean momentum $k_m$ can be
calculated for the transmitted packet. Results for several potential
conditions are shown in Fig. \ref{fig:km}. For ``high'' barriers, 
a wider barrier gives a larger mean momentum in the region $d:[0,4]$. 
This is because as $d$ grows the $|T_k|$ support shifts to 
the higher momentum side.  
Therefore the statement ``Higher momentum components of the packet 
preferably go through the barrier'' applies indeed.
This kind of ``acceleration'' effect is found in other areas of physics 
\cite{obe}.\par
\begin{figure}[htb]
\begin{center}
 \epsfxsize=122mm
 \epsfysize=122mm
  \leavevmode
\epsfbox{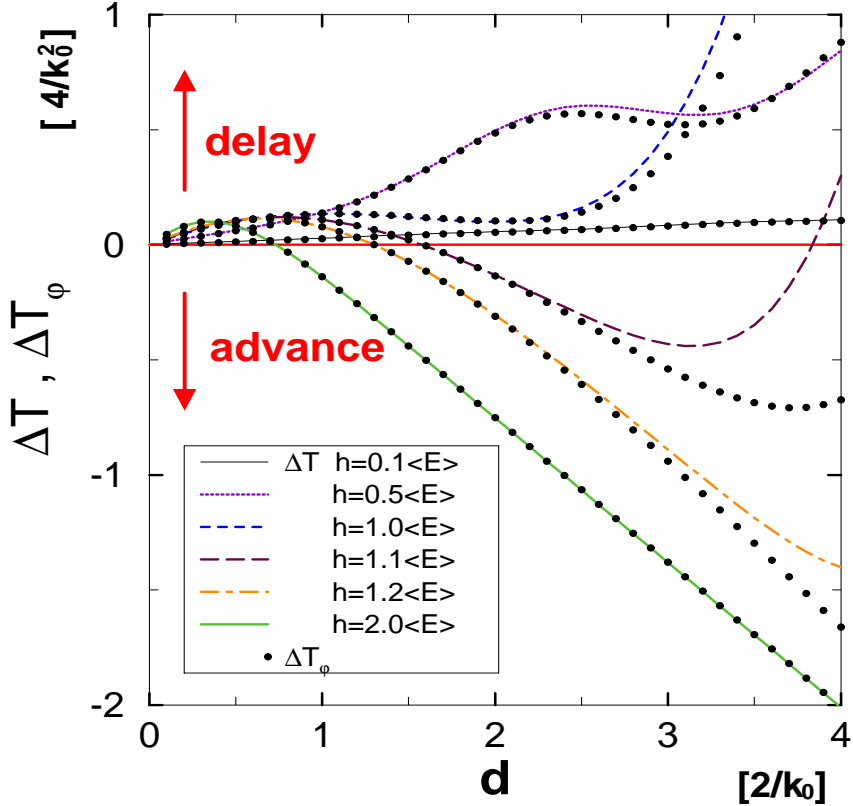}
\vspace{-5mm}
 \caption{The difference between mean arrival times at the barrier exit
 $X=d$ for the tunneling packet and the free one:
$\Delta T\equiv\left\langle T\right\rangle _{d}^{\rm tunnel}
-\left\langle T\right\rangle _{d}^{\rm free}$. $\Delta T _{\varphi}$
(shown by dots) are the same quantity calculated by the stationary
phase method 
for each potential.}
 \label{fig:deltat}
\end{center}
\end{figure}

\begin{quote}
 {\bf Analysis 3: Detection at the barrier exit $X=d$}
\end{quote}
\quad
To see the in-barrier effects more definitely, 
we calculate the difference between mean arrival times 
for the tunneling packet and the free one at the barrier exit $X=d$,
\begin{eqnarray}
\Delta T&\equiv&\left\langle T\right\rangle _{d}^{\rm tunnel}
-\left\langle T\right\rangle _{d}^{\rm free}.
\end{eqnarray}
Results for the same potential
conditions as in Fig. \ref{fig:km} are shown in Fig. \ref{fig:deltat}.
At first we see that in the small $d$ region $\Delta T$ is positive, 
that is, the tunneling packet gets behind the free one, 
for any potential height. However, as $d$ increases, 
$\Delta T$ shows different behaviors according to the potential height.
\par
Roughly speaking, for a ``low'' barrier $\Delta T$ almost stays positive 
but for a ``high'' barrier $\Delta T$ becomes negative.
The ``high'' barrier means that the tunneling modes dominate in the
transmitted packet. In the large $d$ region, $\Delta T$ is negative, 
that is, the tunneling packet goes ahead of the free one 
for the tunneling dominated case. 
We also see a strange behavior where $\Delta T$ changes sign twice 
and finally becomes positive. 
The typical case in Fig. \ref{fig:deltat} is 
the $h=1.1\langle E\,\rangle$ barrier. 
We understand this effect as follows. For a very wide barrier, 
over-the-barrier modes dominate in thetransmitted packet 
($\omega _m=k_m^2/2>h$). That is, as in the ``low'' barrier case, 
$\Delta T$ becomes positive again.
\par
In Fig. \ref{fig:deltat} we also plotted an analogous quantity 
$\Delta T_{\varphi}$ calculated by the stationary phase method. 
We define $\Delta T_{\varphi}$ as follows:
\begin{eqnarray}
\tau _{\varphi}&\equiv
&\left.\frac{d\theta}{d\omega}\right|_{\omega=\omega_m},\\
&&\nonumber\\
\Delta T_{\varphi}&\equiv&\left
( \frac{1}{v_g(k_m)}-\frac{1}{v_g(k_0)}\right)(d-x_0)+\tau _{\varphi},
\end{eqnarray}
where $\theta$ is the phase shift of the transmitted wave defined in
Eq. (\ref {(27)}) and $v_g(\cdot)$ are the group velocities
$v_g(k_m)=\left.(d\omega/dk)\right|_{k=k_m}=k_m$, $v_g(k_0)=k_0$. 
In the ordinary tunneling time problem context $\tau _{\varphi}$ 
is called the phase time. As seen in Fig. \ref{fig:deltat},
although $\Delta T_{\varphi}$ has good agreement with our $\Delta T$ 
in the small $d$ region, as $d$ increases, the difference becomes clear 
for $h\simeq\langle E\,\rangle$ barriers.
This is because for such barriers the momentum distribution
$e^{-\sigma ^2 (k-k_0)^2/2}|T_k|$ is no longer symmetric 
with respect to $k_m$, and the packet's peak given by 
the stationary phase method loses physical significance.
\par
%\vspace{7mm}
We would like to close this section by referring to the relationship 
between our results and the experiment by Chiao and co-workers, 
that is, tunneling of the massless photon.
Of course our model does not describe the propagation of photons, 
and we now mention only the qualitative behavior. 
Because the energy of the photon in the vacuum is 
exactly proportional to its momentum, 
the group velocity of the photon after tunneling is  constant $c$. 
Therefore we get an $X$ independent constant value of the difference  
$\Delta T=\left\langle T\right\rangle _{X}^{\rm tunnel}
-\left\langle T\right\rangle _{X}^{\rm free}$ at any $X\geq d$. 
Since their experimental setup is the tunneling dominated one,   
it may correspond to our model with high and medium wide
barriers. Then our results are consistent with their experimental
observation that the tunneling photon arrives earlier than the free photon. 
%%%%%%%%%%%%%%%%%%%%%%%% Section 4 %%%%%%%%%%%%%%%%%%%%%%%%%%%%% 
\section{Nelson's stochastic interpretation}
\quad
Now we consider the stochastic interpretation of quantum mechanics
introduced by Nelson. 
This approach interprets the motion of particles in quantum mechanics 
as ``real-time'' stochastic processes \cite{nel}.
Nelson substituted the coordinate variable $x(t)$ 
for a stochastic variable performing the Brownian motion 
in a certain drift force field.
The time evolution of $x(t)$ is described by the Ito-type stochastic
differential equation,
\begin{equation}
 dx(t)=b\left(x(t),t\right)dt+dw(t),\label{(7)}
\end{equation}
where $b(x,t)$ is the so-called drift term,
given by the ordinary Schr\"odinger wave function $\psi (x,t)$ as
\begin{equation}
 b(x,t)=\frac{\hbar}{m}\frac{\partial}{\partial x}
         \left({\rm Im+Re}\right){\rm ln}\psi (x,t). \label{(8)}
\end{equation} 
The Gaussian noise $dw$ characterizes the stochastic behavior
and should have the following statistical properties:
\begin{eqnarray}
   \langle dw(t)\rangle=0, \quad 
   \langle dw(t)dw(t)\rangle=\frac{\hbar}{m}dt.\label{(10)}
\end{eqnarray}
Starting with an initial distribution of $x(0)$ we solve Eq. (\ref {(7)})
and obtain sample paths. 
Averaging a physical variable with these sample paths, 
we can calculate the expectation value 
for the ordinary probability distribution $\left|\psi
(x,t)\right|^2$. In this approach,
we are able to observe ``trajectories'' of real-time motion of a particle, 
that is, to describe the quantum mechanical time evolution 
by a classical stochastic process. 
\par
\begin{figure}[htb]  
\hspace{0mm}
\vspace{0mm} 
 \parbox{70mm}{
 \epsfxsize=70mm     
 \epsfysize=70mm
  \leavevmode
\epsfbox{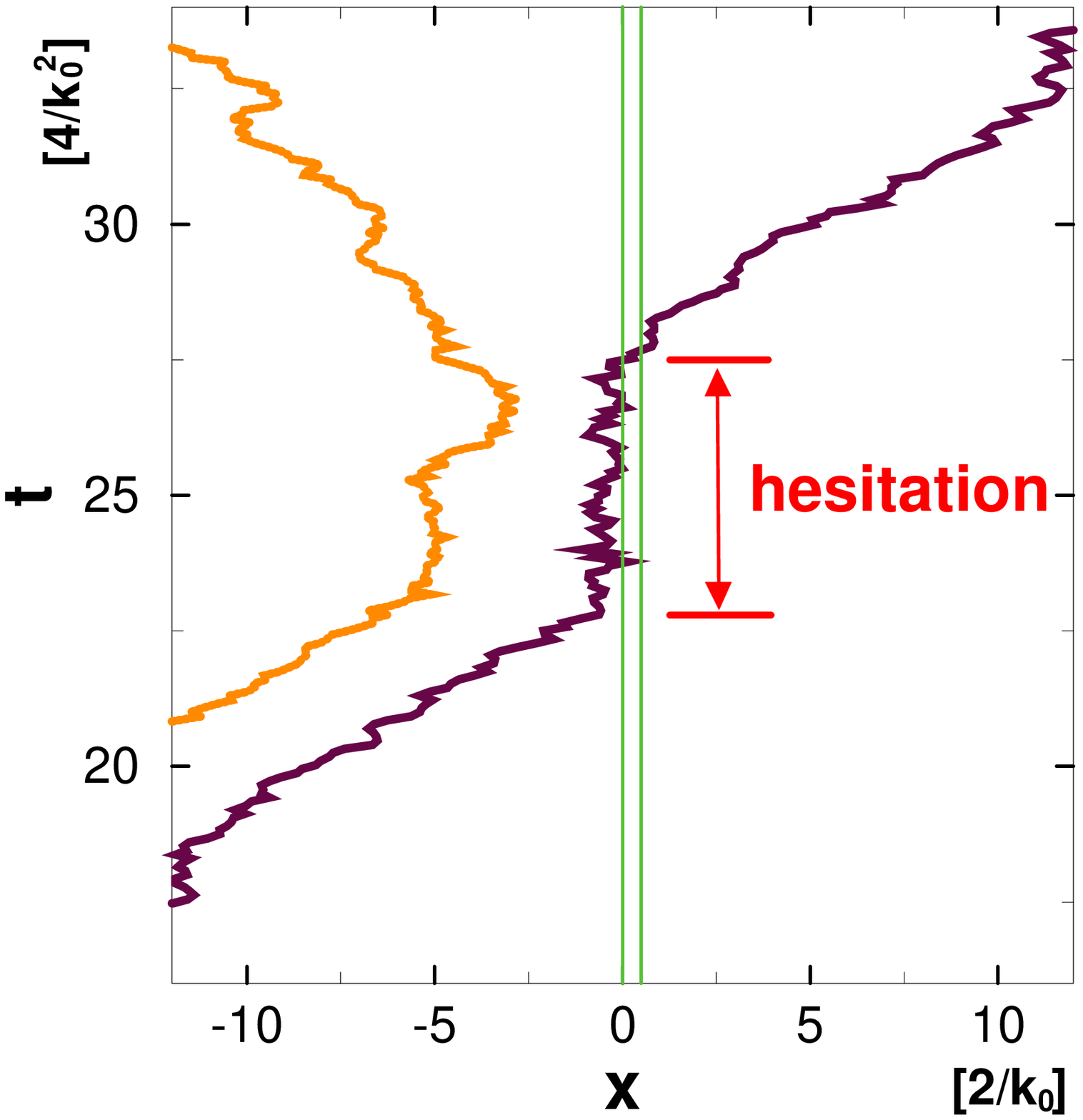}
\vspace{-5mm}
 \caption{Typical sample paths with hesitation}
 \label{fig:paths}     
 }
\hspace{2mm} 
\parbox{70mm}{
 \epsfxsize=70mm      
 \epsfysize=70mm
 \leavevmode
\epsfbox{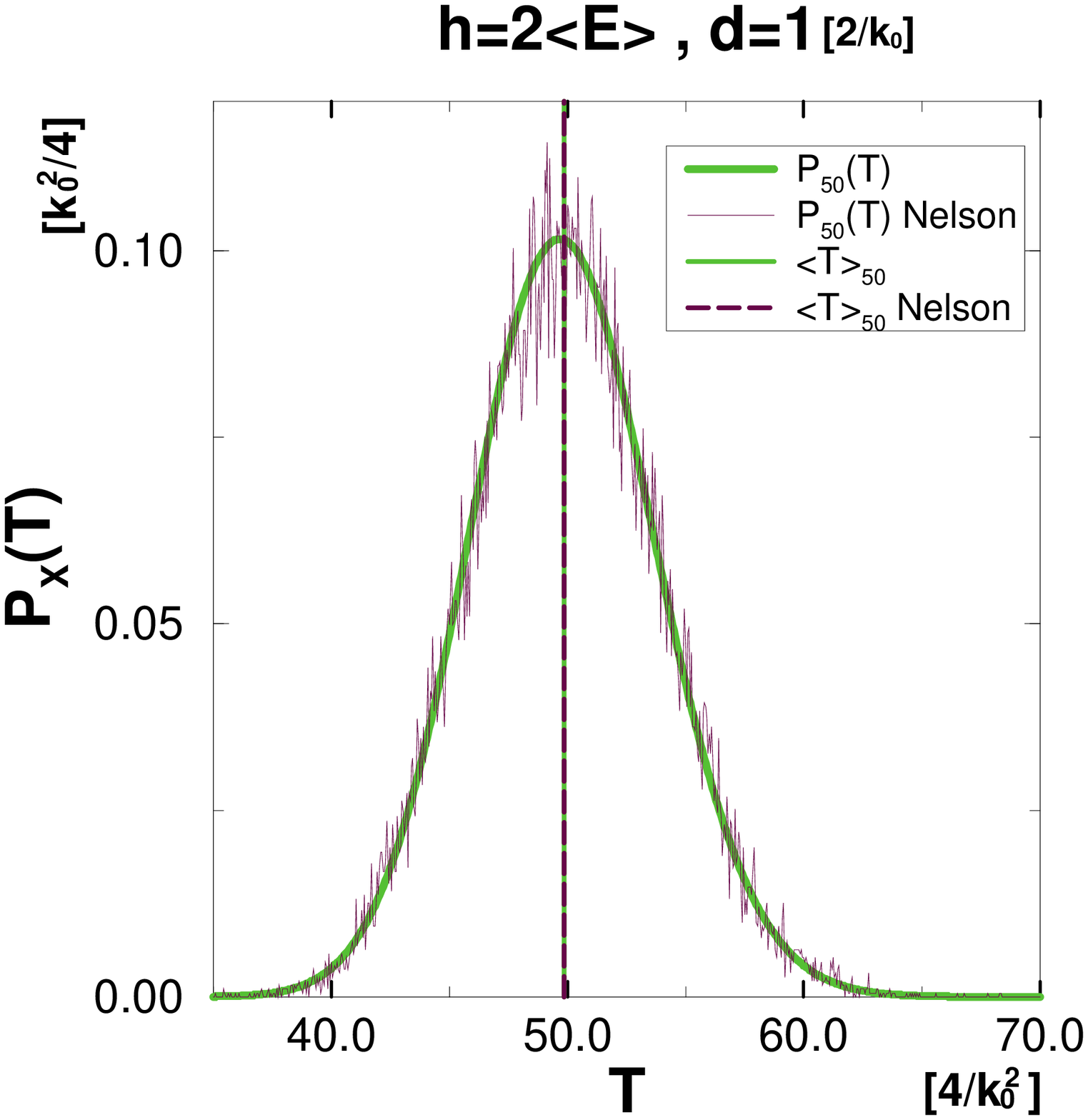}
\vspace{-5mm}
 \caption{Comparison of two methods~~~~~~}
 \label{fig:neldis} 
 }
\end{figure}

 Thus in Nelson's approach 
it may be possible to understand an imaginary-time process 
such as tunneling in real-time language. 
It was pointed out that the tunneling particle ``hesitates'' 
in front of the barrier as seen in Fig. \ref{fig:paths} \cite{ohba}. 
This fact was understood to imply that the particle tunneling 
through the barrier should always be delayed compared with the free one 
because of this hesitation. 
Is it contradictory to our results? 
Nelson's approach can reproduce physical quantities 
in standard quantum mechanics, and there cannot be any conflict.
\par

Now we analyze the mean arrival time in Nelson's stochastic interpretation. 
One intuitive idea of defining the arrival time for a sample path is to
measure the time for a path to reach a detecting point for the first time:
``the first time counting scheme'' \cite{hasi}. 
However, this notion has no counterpart in the physical quantities 
of standard quantum mechanics.  
We have to work with the probability of existence of paths at a point
(or a section) at some definite time. 
The difference between these two notions is that the latter counts 
the possibility of a path going beyond the point 
and coming back to it at the measuring time.
\par 
We define a probability function $\rho_{X}^{\rm N}(T)$, 
\begin{eqnarray}
\rho_X^{\rm N}(T)~dx=\frac{n(X,T)}{N},\label{(12)}
\end{eqnarray}
where $N$ is the total number of sample paths 
and $n(X,T)$ is the number of sample paths that exist in $[X,X+dx]$ 
at time $T$. As stressed before, 
we will count the number of paths passing a target point 
over and over again, i.e., we now employ ``the multiple counting scheme.'' 
With this scheme, 
we define the arrival time distribution $P_{X}^{\rm N}(T)$
and the mean arrival time $\left\langle T\right\rangle _{X}^{\rm N}$ 
of the particle in Nelson's stochastic interpretation,
\begin{eqnarray}
 &&P_{X}^{\rm N}(T)~dT=
      \frac{\rho_X^{\rm
      N}(T)~dT}{{\displaystyle\int_0^{\infty}dT\rho_X^{\rm N}(T)}}
      ,\label{(13)}\\ 
 &&\left\langle  T\right\rangle _{X}^{\rm N}
  =\int _0^{\infty}T P^{\rm N}_{X}(T)dT.\label{(14)}
\end{eqnarray}
\par
We calculate $P_X^{\rm N}(T)$ and $\left\langle T\right\rangle_X^{\rm N}$ 
with an $h=2\left\langle E\right\rangle$, $d=1$ barrier 
by solving Eq. ({\ref{(7)}}) to get $N=10^6$ sample paths.
The result is shown in Fig. \ref{fig:neldis}.
The distribution $P_X^{\rm N}(T)$ agrees with $P_X(T)$ very well; 
therefore $\left\langle T\right\rangle _X^{\rm N}$ agrees with 
$\left\langle T\right\rangle _X$. 
The distribution given by Nelson's approach exactly reproduces 
our previous results, just as expected. 
Of course, if we employ the first time counting scheme, 
$P_X^{\rm N}(T)$ shifts to an earlier time region 
and therefore $\left\langle T\right\rangle _X^{\rm N}$ is smaller 
than $\left\langle T\right\rangle _X$.
\par
\begin{figure}[htb]  
\hspace{0mm}
 \parbox{70mm}{
 \epsfxsize=70mm     
 \epsfysize=70mm
  \leavevmode
\epsfbox{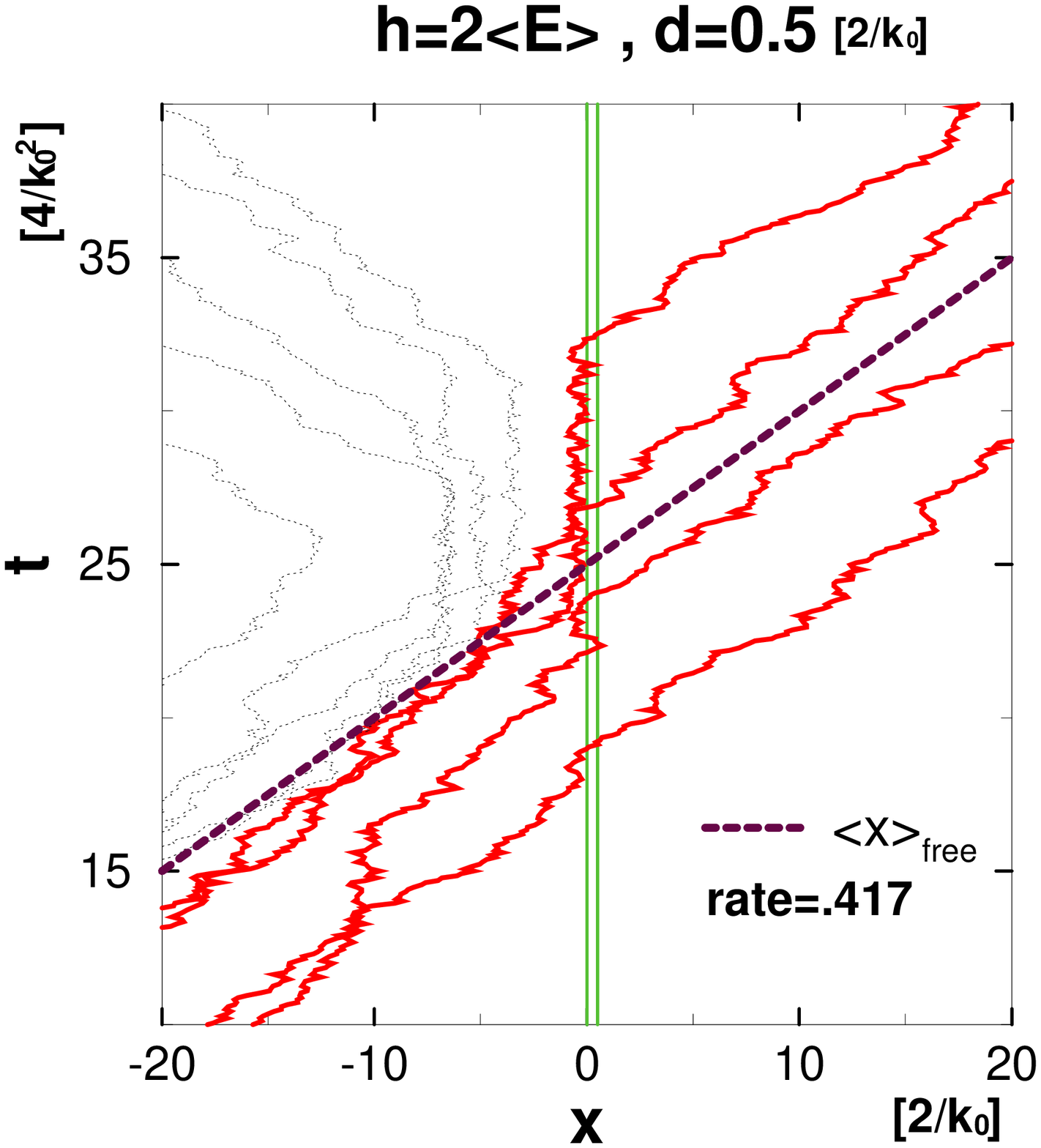}
\vspace{-5mm}
 \caption{Sample paths for $\Delta T>0$}
 \label{fig:path05}     
 }
\hspace{2mm} 
\parbox{70mm}{
 \epsfxsize=70mm      
 \epsfysize=70mm
 \leavevmode
\epsfbox{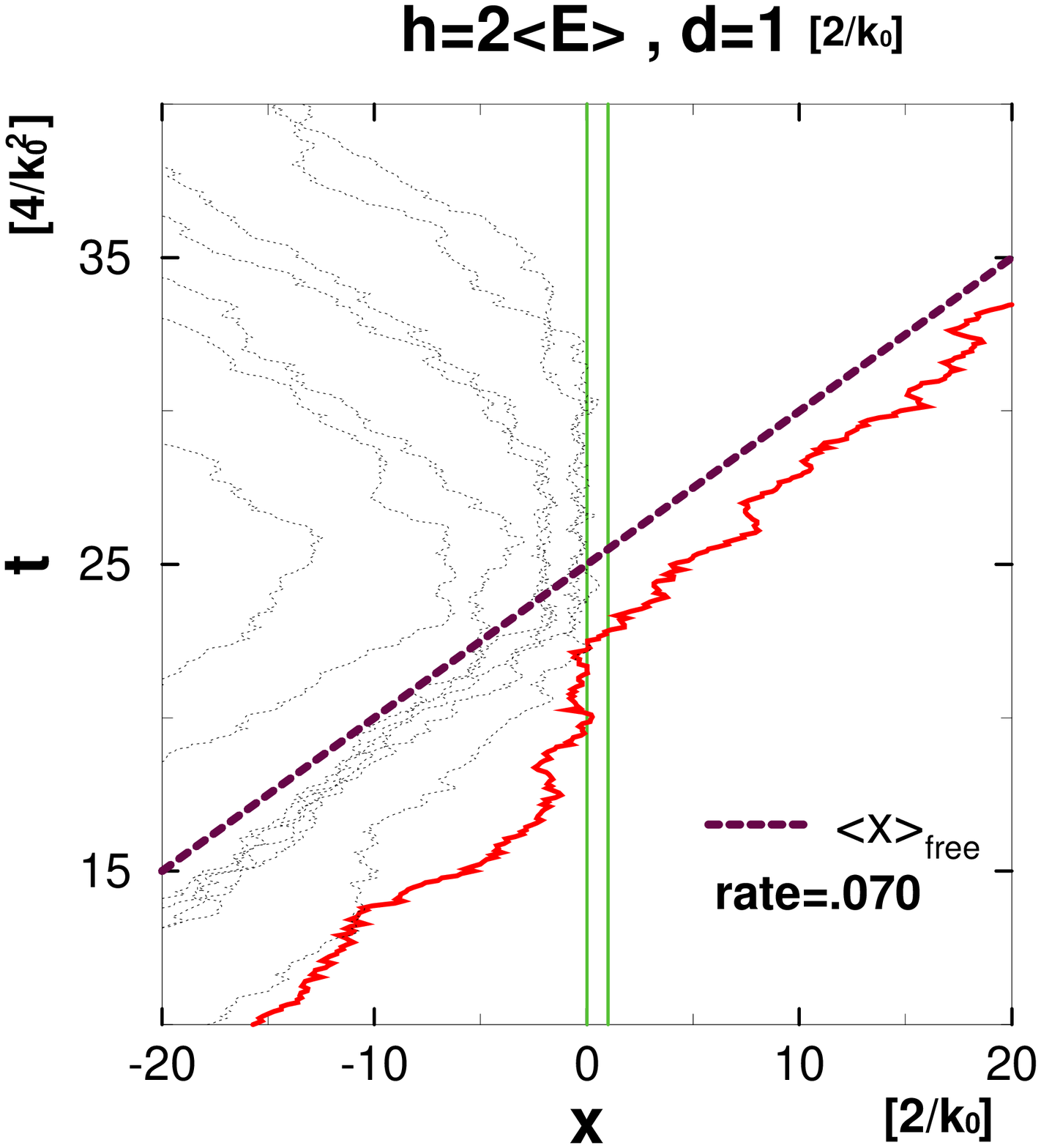}
\vspace{-5mm}
 \caption{Sample paths for $\Delta T<0$}
 \label{fig:path10} 
 }
\end{figure}
\par
%\vspace{-3mm}
\vspace{0mm}
\begin{figure}[htb]  
\hspace{0mm}
 \parbox{70mm}{
 \epsfxsize=70mm     
 \epsfysize=70mm
  \leavevmode
\epsfbox{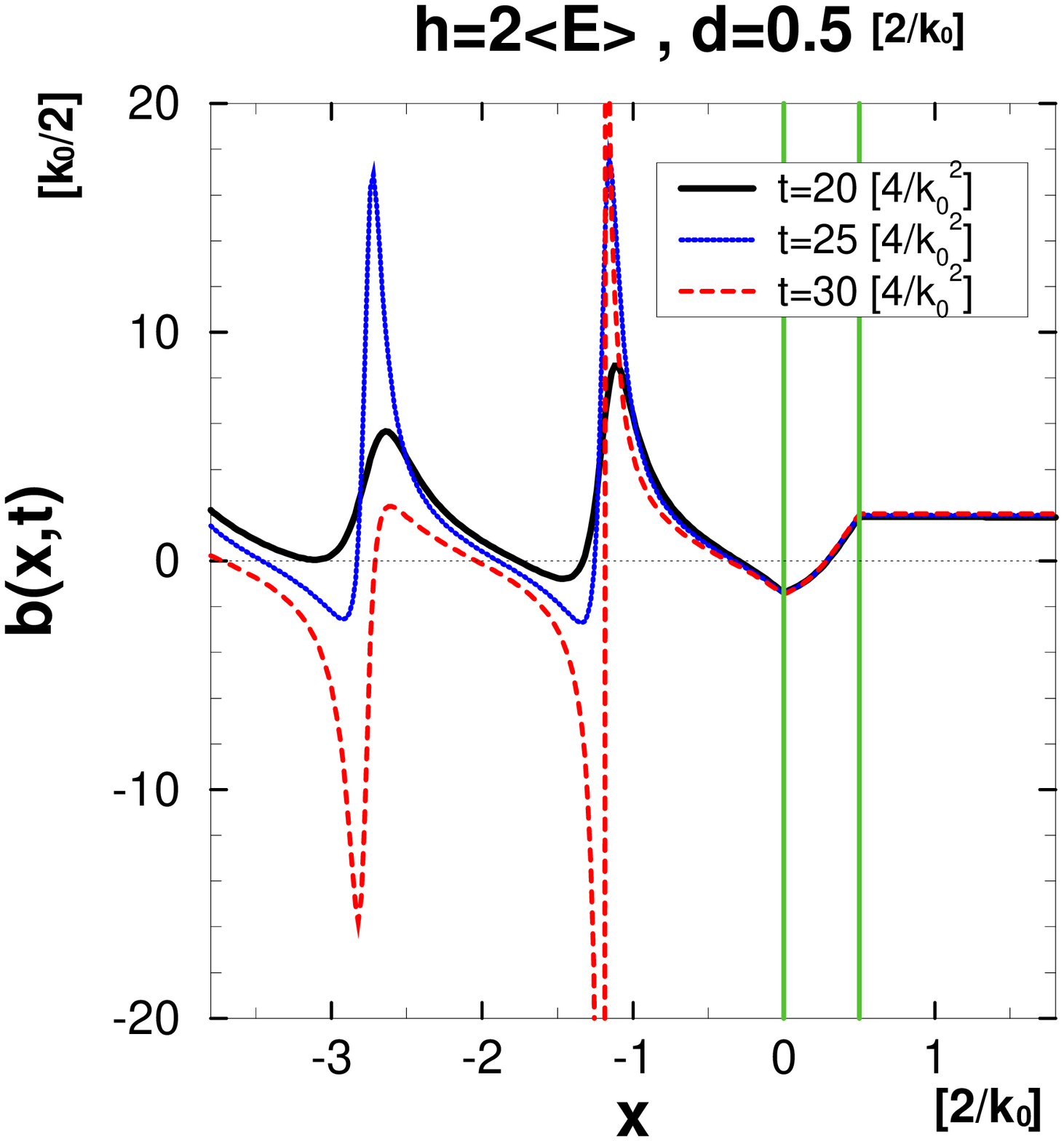}
\vspace{-5mm}
 \caption{Drift velocity for $\Delta T>0$}
 \label{fig:drift05}     
 }
\hspace{2mm} 
\parbox{70mm}{
 \epsfxsize=70mm      
 \epsfysize=70mm
 \leavevmode
\epsfbox{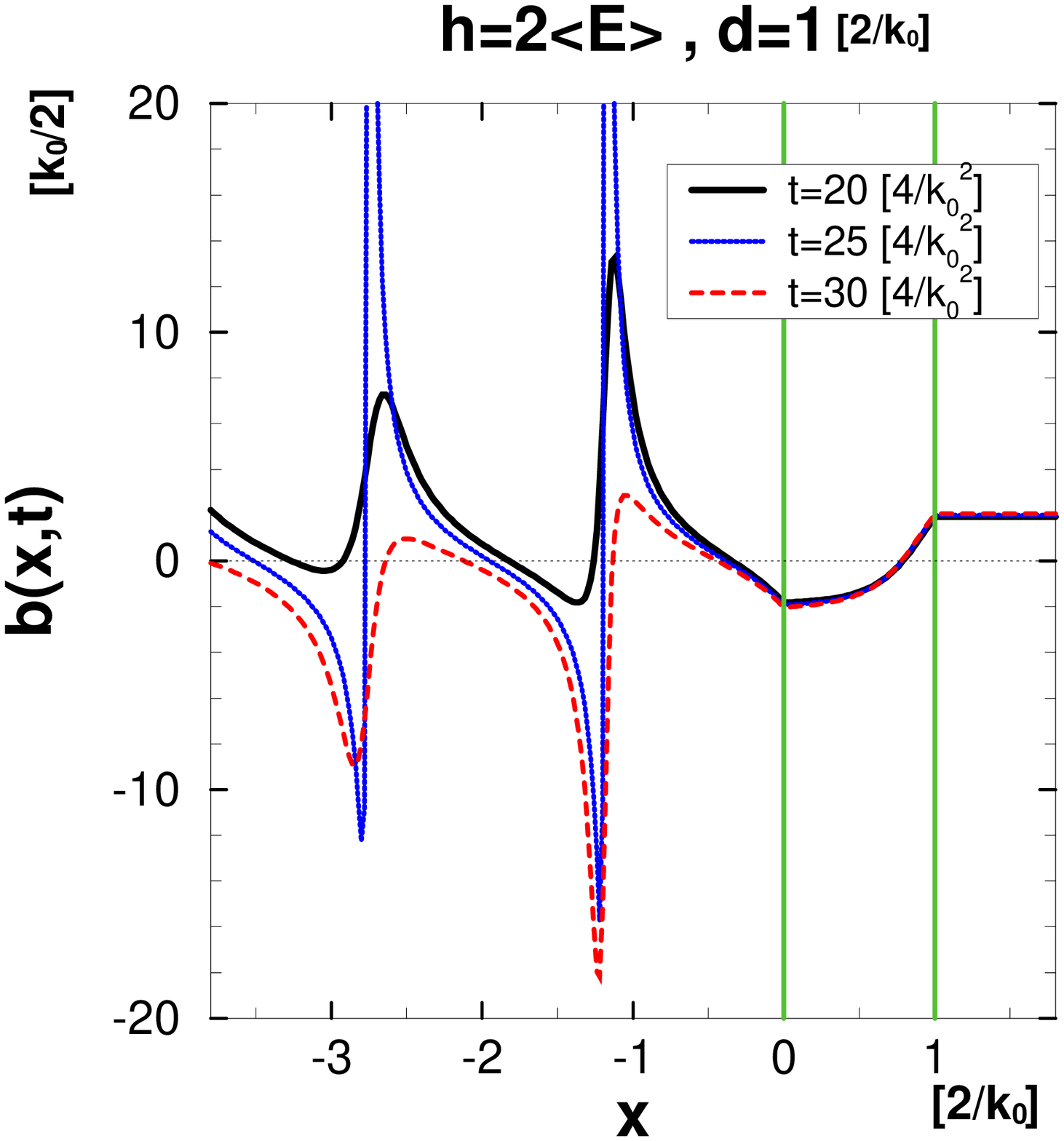}
\vspace{-5mm}
 \caption{Drift velocity for $\Delta T<0$}
 \label{fig:drift10} 
 }
\end{figure}

Then, we should answer the paradoxical question, 
``Why does a hesitating particle arrive earlier than the free one?'' 
To answer this question, let us compare the two cases of
$\Delta T>0$ and $\Delta T<0$.  
First we show typical sample paths for two cases, 
$\Delta T>0~$($\Delta T=0.084$), $h=2\left\langle E\right\rangle$, 
$d=0.5$ in Fig. \ref{fig:path05} and $\Delta T<0~$($\Delta T=-0.138$), 
$h=2\left\langle E\right\rangle$, $d=1$ in Fig. \ref{fig:path10}. 
In both figures we also plot the average position 
of the free sample paths $\left\langle x\right\rangle_{\rm free}$.
We should pay attention to the point $(x,t)=(0,25)$ 
because $\left\langle x\right\rangle_{\rm free}$ arrives in $x=0$ at $t=25$. 
The two figures, Fig. \ref{fig:path05} and Fig. \ref{fig:path10}, 
make a remarkable contrast. 
That is, although in Fig. \ref{fig:path05} even the paths 
that arrive at $x=0$ later than $t=25$ can pass through the barrier, 
in Fig. \ref{fig:path10} essentially only the paths 
that arrived at $x=0$ earlier than $t=25$ can go through it. 
The ``hesitation'' property is seen in both cases. 
In Fig. \ref{fig:path10}, however, even with ``hesitation,'' 
the averaged tunneling path can appear at the barrier exit $x=d$ 
earlier than the averaged free path 
because the tunneling path arrived at $x=0$ much earlier 
than $\left\langle x\right\rangle_{\rm free}$. 
This is the key to the mystery between hesitation and advancement.
\par Well, 
why do the tunneling paths conduct themselves in such a strange way? 
In the first place, why does hesitation occur?
The reason is hidden in the time dependence of the drift velocity
$b(x,t)$. We show $b(x,t)$ for the same conditions discussed above,
especially near the potential barrier 
(Fig. \ref{fig:drift05} and Fig. \ref{fig:drift10}).
In the foreground of the barrier, according to the interference 
of the incident packet and the reflected packet, 
$b(x,t)$ oscillates frequently and becomes null many times. 
Especially near the barrier entrance $x=0$, 
$b(x,t)$ changes from positive to negative, where the particle  
is ``trapped.'' 
These effects cause the path's hesitation. 
\par At earlier times, 
$b(x,t)$ is almost always positive value but at later times, 
it becomes almost always negative. In Fig. \ref{fig:drift10}, 
this tendency is extreme and realization of the tunneling path is 
much rarer than in Fig. \ref{fig:drift05}. 
This is the reason why the early arrived paths tend to pass the barrier 
more easily. 
After all, there is no inconsistency between our results 
and the hesitation behavior in Nelson's interpretation.
\par 
Furthermore, Nelson's interpretation provides us 
an intuitive explanation of our results. 
Let us consider the high potential barrier case. 
It is important that every transmitted path hesitates to some extent. 
In the small $d$ region, because of the high transmission rate, 
even a path arriving at $x\simeq 0$ relatively late can pass the barrier, 
and as a result we find $\Delta T>0$. As $d$ increases, 
the transmission rate becomes lower and only the paths arriving 
at $x\simeq 0$ earlier can penetrate the barrier, 
and as a result we find $\Delta T<0$. Finally, as $d$ becomes very large, 
the paths arriving at $x\simeq 0$ very early hesitate 
there for a very long time;
 therefore $\Delta T$ becomes positive again.
\par
Of course, we must recall that the ``path'' in Nelson's view 
never corresponds to a real particle in ordinary quantum mechanics,  
and the explanation we gave above is just an interpretation. 
The same is true for an interpretation by the Bohm trajectory 
\cite{leav,ml}.
It may be interesting to regard the path as a physical one 
and to calculate various quantities that cannot be calculated 
in ordinary quantum mechanics (the tunneling time 
$\Delta^{\rm N}=\left\langle T\right\rangle _{d}^{\rm N}
-\left\langle T\right\rangle _{0}^{\rm N}$, 
quantities calculated in the first time counting scheme, etc.). 
Although these attempts may give us deeper insights into quantum dynamics,
the validity and significance of them have not been argued much so far 
\cite{ohba2,hino}.
%%%%%%%%%%%%%%%%%%%%%%%% Section 5 %%%%%%%%%%%%%%%%%%%%%%%%%%%%% 
\section{Summary}
\quad Supposing an ideal detector, 
we defined simple expressions for the arrival time distribution $P_X$ 
and the mean arrival time $\left\langle T\right\rangle_X$, 
and applied them to analysis of wave packet tunneling. 
We defined $\Delta T\equiv\left\langle T\right\rangle _{d}^{\rm tunnel}
-\left\langle T\right\rangle _{d}^{\rm free}$ and calculated it for
various barrier conditions and showed the barrier effects clearly.  
In the small $d$ region, $\Delta T$ is always positive, 
but as $d$ increases, $\Delta T >0$ for the over-the-barrier case 
and $\Delta T <0$ for the tunneling case. 
After tunneling, the packet usually moves faster than the free one 
because it preferentially consists of the higher momentum modes 
of the incident packet.
The barrier works as an acceleration filter in a sense.
We also clarified that the stationary phase method gives a good
approximation to our results, particularly in the small $d$ region.
\par
We also confirmed that the stochastic interpretation 
introduced by Nelson reproduces our results. 
Furthermore, we clarified how the ``hesitation'' of the tunneling paths 
in Nelson's picture is consistent with the advancement 
of the tunneling packet.
The key observation is that the paths arriving at the barrier earlier 
than the free mean paths tend to penetrate the barrier more easily.
We pointed out that this property can be explained by the time dependence 
of the drift velocity $b(x,t)$ and found that the behavior of $\Delta T$ 
is intuitively understandable with Nelson's language.
%\thanks
\section*{Acknowledgments}
\pagestyle{myheadings}                      
\quad 
We would like to thank T. Hashimoto, E. M. Ilgenfritz, K. Imafuku, 
K. Morikawa, I. Ohba, T. Tanizawa, H. Terao, and M. Ueda 
for fruitful and encouraging discussions and suggestions. 
We are also grateful to J. G. Muga for comments and for 
calling our attention to some relevant references.
%%%%%%%%%%%%%%%%%%%%%%% References  %%%%%%%%%%%%%%%%%%%%%%%%%%%%% 

\end{document}